\documentclass [10pt]{article}
\title{Non-time-orthogonal reference frames\\in the theory of relativity}
\author{Robert D. Klauber\\1100 University Manor, 38B, Fairfield, Iowa 52556\\email  rklauber@netscape.net}
\date{May 10, 2000}
\usepackage {graphicx}
\usepackage {longtable}

\begin{document}
\maketitle

\bigskip
\begin{abstract}
A simple, though rarely considered, thought experiment on relativistic 
rotation is described in which internal inconsistencies in the theory of 
relativity seem to arise. These apparent inconsistencies are resolved by 
appropriate insight into the nature, and unique properties, of the 
non-time-orthogonal rotating frame. The analysis also explains a heretofore 
inexplicable experimental result.
\end{abstract}

\section*{I. Introduction}

Although addressed by Einstein\footnote{ Albert Einstein, ``Die Grundlage 
der allgemeinen Relativitätstheorie,'' \textit{Ann. Phys.(Leipzig) 
}\textbf{49}, 769-822 (1916).} $^{,}$\footnote{ Albert Einstein, \textit{The 
Evolution of Physics,} (Simon-Schuster, New York, 1938), pp. 226-234.} 
$^{,}$\footnote{ John Stachel, "Einstein and the Rigidly Rotating Disk," 
Chap. 1 in H. Held, \textit{General Relativity and Gravitation} (Plenum, New 
York, 1980), pp. 1-15.} and others\footnote{ P. Ehrenfest, "Gleichfömrige 
Rotation starrer Körper und Relativitätheorie," \textit{Phys. Z.} 
\textbf{10}, 918-928 (1909).} $^{,}$\footnote{ Philip Franklin., "The 
meaning of rotation in the special theory of relativity", \textit{Proc. Nat. 
Acad. Sci. USA} \textbf{8}(9)\textbf{,} 265-268 (1922).} $^{,}$\footnote{ 
M.G. Trocheries, "Electrodynamics in a rotating frame of reference," 
\textit{Phil. Mag.} \textbf{40}(310), 1143-1154 (1949).} $^{,}$\footnote{ 
Hyoitiro Takeno, "On relativistic theory of rotating disk", \textit{Prog. 
Theor. Phys. }\textbf{7}(4), 367-376 (1952).} in the first half of the 
twentieth century, the relativistically rotating reference frame continues 
to be a research topic of interest, and, in fact, has often generated 
significant discussion and debate.\footnote{ Oyvind Gron., "Rotating frames 
in special relativity analyzed in light of a recent article by M. 
Strauss"\textit{,} \textit{Int. J. Theor. Phys.} \textbf{16}(8), 603-614 
(1977).} $^{,}$\footnote{ Nicholas Sama, "On the Ehrenfest paradox," 
\textit{Am. J. of Phys. }\textbf{40}, 415-418 (1972).} $^{,}$\footnote{ 
Gerald N. Pellegrini and Arthur R. Swift, "Maxwell's equations in a rotating 
medium: Is there a problem?\textit{," Am. J. Phys.} \textbf{63}(8), 694-705 
(1995).} $^{,}$\footnote{ Thomas A. Weber, ``Measurements on a rotating 
frame in relativity, and the Wilson and Wilson experiment,'' \textit{Am. J. 
Phys.}\textbf{ 65} , 946-953 (1997).} $^{,}$\footnote{ Charles T. Ridgely., 
``Applying relativistic electrodynamics to a rotating material medium'', 
\textit{Am. J. Phys}. \textbf{66} (2) 114-121 (1998).} $^{,}$\footnote{ 
Robert D. Klauber, ``Comments regarding recent articles on relativistically 
rotating frames'', \textit{Am. J. Phys.} \textbf{67}(2), 158-159, (1999).} 
$^{,}$\footnote{ Thomas A. Weber, ``Response to `Comments regarding recent 
articles on relativistically rotating frame' [Am. J. Phys. 67(2), 158 
(1999)'', \textit{Am. J. Phys. }\textbf{67}(2), 159-160 (1999).} 
$^{,}$\footnote{ Robert V. Krotkov, Gerald N. Pellegrini, Norman C. Ford, 
and Arthur R. Swift, "Relativity and the electric dipole moment of a moving, 
conducting, magnetized sphere\textit{," Am. J. Phys.} \textbf{67}(6), 
493-498 (1999).} $^{,}$\footnote{ Hrvoje Nikolic, ``Ehrenfest paradox, 
non-time-orthogonal frames, and local observers,'' \textit{Los Alamos Nat 
Lab}, xxx.lanl.gov, paper num gr-qc/9904078 (1999).} (References cited in 
this section are not exhaustive.) Some\footnote{ Franco Selleri, 
``Noninvariant one-way speed of light and locally equivalent reference 
frames,'' \textit{Found. Phys. Lett.}, \textbf{10}, 73-83 (1997)\textbf{.}} 
$^{,}$\footnote{ J. Paul Wesley, \textit{Classical Quantum Theory }(Benjamin 
Wesley, Blumberg, Germany, 1996), pp 202-203.} $^{,}$\footnote{ R. Anderson, 
I. Vetharaniam, and G. E. Stedman, ``Conventionality of Synchronisation, 
Gauge Dependency, and Test Theories of Relativity,'' \textit{Phys. Rep.}, 
\textbf{295} (3\&4), 93-180 (1998).} , who have considered the thought 
experiment described below and/or the Sagnac experiment\footnote{ G. Sagnac, 
\textit{Comptes Rendus, }\textbf{157}, 708 (1913).} $^{,}$\footnote{ E.J. 
Post, "Sagnac effect," \textit{Mod. Phys}. \textbf{39}, 475-493 (1967). } , 
have proposed the possible existence of a preferred reference frame, which 
is somehow disguised within our present understanding of special relativity.

The present article does not support this view, and in the final analysis, 
is consonant with the special and general theories of relativity. It does, 
however, illustrate one way in which the extant theory does appear to be 
self-contradictory. This seeming internal incongruity is resolved by 
analysis of the non-time-orthogonal nature (i.e., time is not orthogonal to 
at least one spatial dimension) of the rotating frame. In the process, 
however, non-time-orthogonal (NTO) frames are found to exhibit certain 
unique and somewhat surprising characteristics. Though some of these 
characteristics are not what might presently be considered typically 
relativistic, they are not only in agreement with empirical evidence, but 
explain what has heretofore been considered an anomalous experimental 
result.

\section*{II. Rotating Frames in Thought and Practice}

\subsection*{A. Simple Gedanken Experiment}

Consider the rotating disk of Figure 1 with a rim mounted light source 
capable of emitting light pulses in both directions along the disk 
circumference. A cylindrical mirror, polished side facing inward, is mounted 
on the rim as well.

\begin{figure}[htbp]
\centering
\includegraphics*[bbllx=0.26in,bblly=0.11in,bburx=5.11in,bbury=1.84in,scale=1.00]{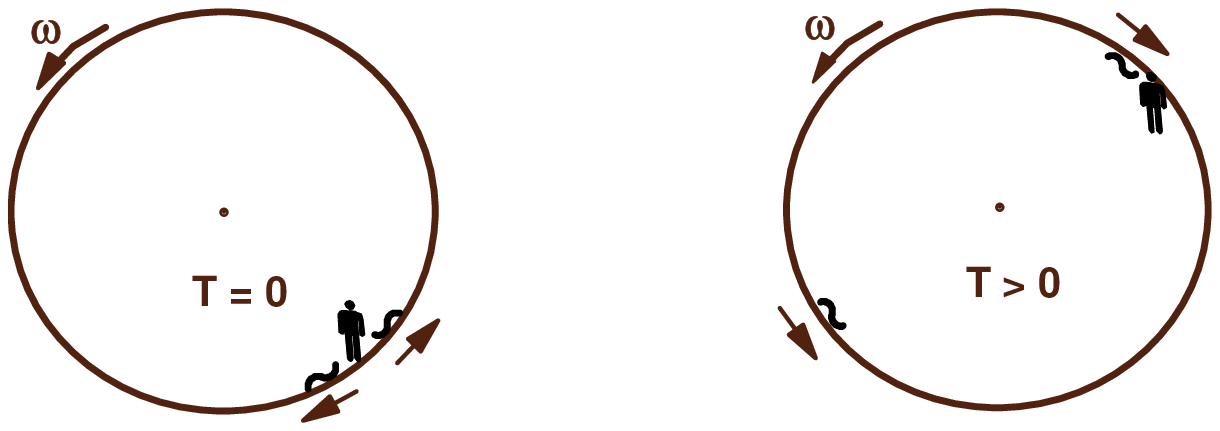}
\end{figure}

\bigskip

\bigskip

\begin{center}
\textbf{Figure 1. Thought Experiment}
\end{center}

\bigskip

At time \textit{T = }0, an observer attached to the light source triggers 
the emission of two very short light pulses, one in the clockwise direction, 
the other in the counter clockwise direction. Each light wave packet is 
1$^{o}$ long at the rim radius, i.e., it is 1/360 of the rim circumference.

In the lab frame each light pulse has speed \textit{c}, and each is 
reflected by the cylindrical mirror, such that it travels a circular route 
with radius equal to the disk radius \textit{r}. As the two light pulses are 
travelling, the disk is rotating c.c.w. at an $\omega  $high enough to 
produce relativistic rim velocities. Since the observer mounted on the light 
source is moving c.c.w. as well, the c.w. moving light pulse reaches him 
before the c.c.w. pulse does. From the point of view of the lab, this 
conclusion is inescapable, and since we are talking about separate 
detectable physical events, it must be true from the point of view of the 
observer on the disk as well.

The disk observer knows that each light signal traveled the same distance in 
his frame (he set up the experiment). He also knows that one of them took 
less time to travel that distance than the other. Hence, the only conclusion 
he can make is that the speed of light on the disk was greater for the c.w. 
travelling pulse than for the c.c.w. travelling pulse. Due to symmetry, it 
also seems inevitable that this conclusion holds true locally as well as 
globally.

Note this ``experiment'' did not entail any wave interference measurements. 
It dealt simply with arrival times of short photon wave packets and had 
nothing to do with the wave nature of those packets.

The conundrum, of course, is that, according to relativity theory, the speed 
of light is invariant and always equal to \textit{c} in all directions, no 
matter what frame one is in. (Those who believe that the general theory of 
relativity may say otherwise are referred to the Appendix.) The problem is 
compounded when one considers that most analyses of relativistic rotation 
utilize an infinite series of local Lorentz frames instantaneously co-moving 
with the rotating frame at a given radius from the center of rotation. 
Calculations of things like spatial distance around the circumference are 
then made by summing (i.e., integrating) infinitesimal quantities (e.g., 
\textit{dx}) from all of the local co-moving Lorentz frames.

But in Lorentz frames the speed of light is always \textit{c}. So in effect, 
when one uses such frames one is assuming the local speed of light is 
\textit{c}, isotropic and invariant. But, as we have illustrated in our 
thought experiment, this does not appear to be the case for rotating frames. 
And since the invariance of the speed of light is one postulate upon which 
the theory of relativity rests, one must immediately re-evaluate not only 
the suitability of such approaches, but also the very theory itself.

\subsection*{B. The Sagnac Experiment}

The analysis of the previous section literally reeks of a preferred frame 
(``absolute'' space), and proponents of such a thing commonly cite the 
Sagnac experiment (see Figure 2) as empirical proof\footnote{ See, for 
example, reference 18.} .

In the Sagnac experiment, a light beam is emitted radially from the center 
of a rotating disk and is split by a half-silvered mirror M at radius 
\textit{r}. From there one part of the beam is reflected by mirrors 
appropriately placed on the disk such that it travels in one direction 
effectively around the circumference. The other half of the beam travels the 
same route over the same distance, but in the opposite direction. The beams 
then meet up again and are reflected back to the center where interference 
of the two beams results in a fringing, i.e., a displacement of one light 
wave with respect to the other.

If the speed of light on the disk were invariant, then as the rotational 
velocity of the disk increased, the fringe pattern would remain unchanged, 
similar to what one finds in the Michelson-Morley experiment. However, when 
this was done by Sagnac$^{20}$ and others$^{21}$ who have repeated his 
experiment, the fringe pattern did in fact change, indicating a dependence 
of the rotating frame speed of light on both direction and rotational speed.

\begin{figure}[htbp]
\centering
\includegraphics*[bbllx=0.26in,bblly=0.11in,bburx=4.07in,bbury=3.77in,scale=0.72]{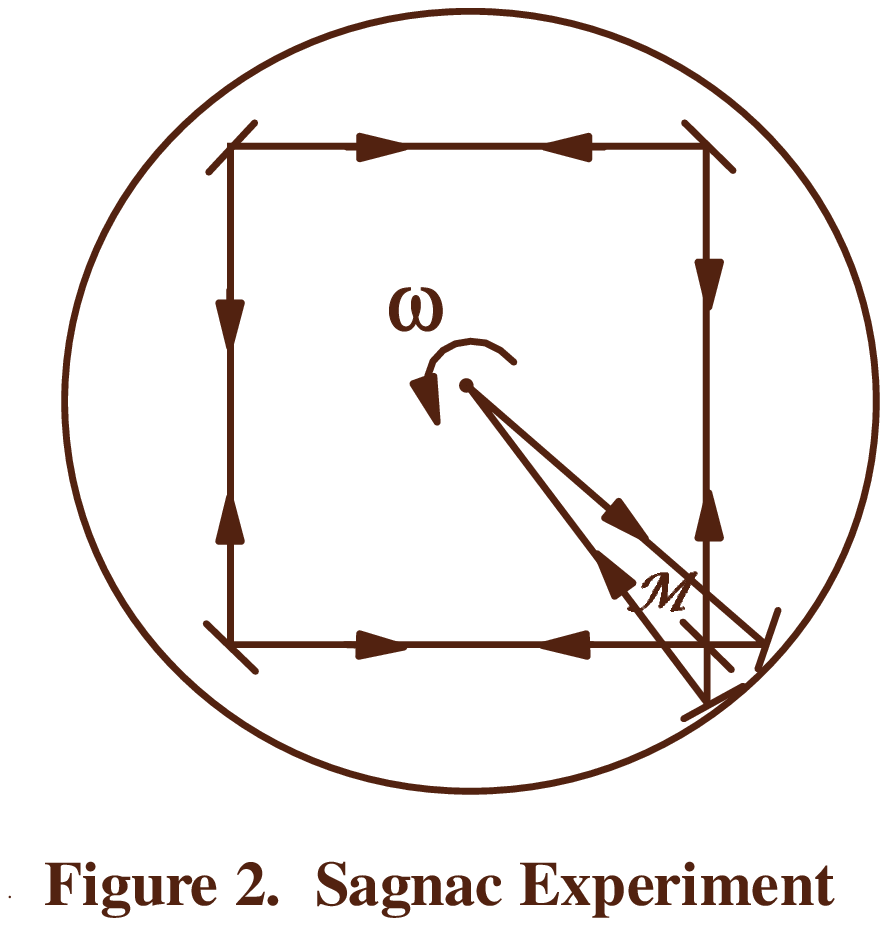}
\end{figure}

\bigskip

The test results have experimental accuracy only to first order in 
\textit{v/c} = $\omega $\textit{r/c}, but they indicate that the speed of a 
light ray tangent to the circumference measured locally on the disk is equal 
to\footnote{ Post (ref. 21) presents the Sagnac results in terms of the 
fringe shift (his equation (1)). Using (1) of the present paper, one can 
derive (1) in Post (use A = $\pi $\textit{r}$^{2}$), and vice versa.}

\begin{equation}
\label{eq1}
|v_{light} |\; \cong \;c \pm \omega r,
\end{equation}

\noindent
where the approximately equal sign implies accuracy to first order, and the 
sign in front of the last term depends on the relative direction of the rim 
tangent and light ray velocities. This same relationship can be easily 
derived using the logic of the previous section. It is obviously 
disconcerting, as (1) looks far more Galilean/Newtonian in nature than 
relativistic.

Some have attributed the Sagnac phenomenon to wave effects. Mashoon et 
al\footnote{ Bahram Mashhoon, Richard Neutze, Mark Hannam, Geoffrey E. 
Stedman, ``Observable frequency shifts via spin-rotation coupling'', 
\textit{Phys. Lett. A, }\textbf{249}, 161-166 (1998).} , for example, 
consider it to be a ``manifestation of the coupling of orbital angular 
momentum of a particle .. to rotation''. For a wave this perturbation in the 
Hamiltonian induces a phase shift such as that measured in the Sagnac 
experiment. In somewhat similar fashion, Anandan\footnote{ J. Anandan, 
``Sagnac effect in relativistic and nonrelativistic physics,'' \textit{Phys. 
Rev. D }\textbf{\textit{24}}(2), 338-346 (1981).} asserts ``.. this effect 
depends only on the frequency of the beams ...''.

However, such analyses fail to answer the question raised by our thought 
experiment, which was not based on wave interference, but solely on arrival 
times of very short wave packets. As (1) is readily deduced from that 
thought experiment, it appears a more fundamental reconciliation with the 
theory of relativity is needed.

\section*{III. Transformation to the Rotating Frame}

We will adopt what is presently the most widely$^{10,11,}$\footnote{ C. 
Moller, \textit{The Theory of Relativity} (Clarendon Oxford, 1969).} 
$^{,}$\footnote{ L. D. Landau, and E. M. Lifshitz, \textit{The Classical 
Theory of Fields} (Addison-Wesley, Reading, 1962), pp. 271-298.} 
$^{,}$\footnote{ Oyvind Gron,., "Relativistic description of a rotating 
disk"\textit{,} \textit{Am. J. of Phys.} \textbf{43}(10), 869-876 (1975).} 
$^{,}$\footnote{ Robert D. Klauber, ``New perspectives on the relatively 
rotating disk and non-time-orthogonal reference frames'', \textit{Found. 
Phys. Lett. }\textbf{11}(5), 405-443 (1998). On page 421 Klauber lists 
assumptions upon which the transformation is based.} $^{,}$\footnote{ Adler, 
R., Bazin, M., and Schiffer, M., \textit{Introduction to General Relativity} 
(McGraw-Hill, New York, 1975), 2$^{nd}$ ed., p. 121-122.} , though not 
universally$^{17,}$\footnote{ Post (reference 21) noted the transformation 
is determinable experimentally only to first order and suggested the 
presence of a factor he termed $\gamma  $on the right hand side of (2.a) 
and in front of the second term on the right hand side of (2.c). He 
considered that this factor could be unity, the Lorentz contraction factor, 
or perhaps something else. Ehrenfest, Franklin, and Trocheries (references 
4,5,6) considered transformations other than (2), which never proved 
completely satisfactory. Although Stedman (reference 19 and in private 
communication) contends there is latitude in the choice of transformation 
due to gauge freedom, he does consider the same transformation to the 
rotating frame used by Selleri (ref. 17).} , accepted transformation (see 
(2.a-d) below) between the lab and rotating frames. This coordinate 
transformation, where upper case coordinates represent the inertial frame K, 
lower case denote the rotating frame k, and the axis of rotation is 
coincident with both the Z and z axes, is

\begin{equation}
\label{eq2}
cT = ct\quad \quad \quad \quad \quad \quad \left( {2.a} \right),
\end{equation}

\[
R\; = \;r\quad \quad \quad \quad \quad \quad \left( {2.b} \right),
\]

\[
\Phi \; = \;\phi \; + \;\omega t\quad \quad \quad \quad \left( {2.c} 
\right),
\]

\[
Z\; = \;z\quad \quad \quad \quad \quad \quad \left( {2.d} \right).
\]

$\omega $ is the angular velocity of the disk, and \textit{t}, the 
coordinate time for the rotating system, is the proper time of a standard 
clock located at the origin of the rotating coordinate frame, i.e., it is 
equivalent to any standard clock at rest in K. Note that \textit{t} is only 
a coordinate. It is merely a label and cannot be expected to equal proper 
time at any given point on the disk (except, of course, at \textit{r}=0).

The metric for the rotating system can be found from the line element for 
the standard cylindrical coordinate system of the Minkowski space K

\begin{equation}
\label{eq3}
ds^{2} = - c^{2}dT^{2} + dR^{2} + R^{2}d\Phi ^{2} + dZ^{2}
\end{equation}

Assuming \textit{ds }is invariant, one can find \textit{dT, dR, d}$\Phi 
$\textit{,} and \textit{dZ} from (2), and insert into (3) to obtain the line 
element, and hence the metric, of the coordinate grid in k.

\begin{equation}
\label{eq4}
ds^{2} = - c^{2}\left( {1 - r^{2}\omega ^{2}/c^{2}} \right)dt^{2} + dr^{2} + 
r^{2}d\phi ^{2} + 2r^{2}\omega d\phi dt + dz^{2}
\end{equation}

\[
 = g_{\alpha \beta } dx^{\alpha }dx^{\beta } \quad 
\]

For a fixed point in the rotating frame (i.e., $dr = d\phi = dz = 0$) with 
$ds^{2} = - c^{2}d\tau ^{2}$inserted into $(4)$, we find the local 
proper time on the disk to be

\begin{equation}
\label{eq5}
d\tau = \left( {1 - r^{2}\omega ^{2}/c^{2}} \right)^{1/2}dt = \left( {1 - 
r^{2}\omega ^{2}/c^{2}} \right)^{1/2}dT.
\end{equation}

That this is the familiar Lorentz time dilation factor (with $v = \omega 
r$), in full accord with numerous cyclotron experiments, supports the 
contention that (2.a-d) is indeed the correct form of the transformation.

\section*{IV. Speed of Light in NTO Frames}

Note that the metric $g_{\alpha \beta } $of $(4)$ is non-diagonal, and 
hence the rotating frame is non-time-orthogonal (NTO). The presence of the 
non-zero line element term in $d\phi dt$indicates that in 4D spacetime, the 
time axis is not orthogonal to the spatial axis for the circumferential 
direction. As we shall see, this has profound implications for measurement 
of the speed of light.

\subsection*{A. Analytical Determination of the Speed of Light}

Consider the path of a photon travelling in the circumferential direction, 
e.g., along the disk rim. For light, \textit{ds} = 0 (see (3)), and for the 
path considered, \textit{dr=dz=}0. Inserting these values in $(4)$, 
solving the resultant quadratic equation for $d\phi $, dividing by 
\textit{dt}, and multiplying by \textit{r, }one obtains a coordinate 
velocity for the photon as seen from the rotating frame

\begin{equation}
\label{eq6}
v_{light,circum,coord} \,\, = \;\;\frac{{rd\phi }}{{dt}}\;\; = \;\; - 
r\omega \pm c.
\end{equation}

To find the physical velocity one would measure with standard rods and 
clocks mounted on the rotating frame, we must use (5) to convert coordinate 
time \textit{dt }in (6) to physical time on standard rotating clocks at 
radius \textit{r}. This yields

\begin{equation}
\label{eq7}
v_{light,circum,phys} \,\; = \;\;\frac{{rd\phi }}{{d\tau }}\;\; = 
\;\,\;\frac{{1}}{{\sqrt {1 - r^{2}\omega ^{2}/c^{2}} }}\frac{{rd\phi 
}}{{dt}}\,\;\, = \;\,\,\frac{{ - r\omega \pm c}}{{\sqrt {1 - v^{2}/c^{2}} 
}},
\end{equation}

\noindent
which is the exact form of the approximate (first order) relationship (1) 
deduced from the Sagnac experiment and our thought experiment.

Note that had we started with a time orthogonal frame, there would have been 
no $d\phi dt$ type off diagonal term in $(4)$, and hence no $r\omega $ 
term in (7). (For example, take the lab frame with $\omega = 0$ in (7). 
Alternatively, set the off diagonal term to zero in $(4)$ and follow the 
steps used to derive (7).)

It is important to note that the non-relativistic looking, Newtonian-like, 
velocity addition relationship of (7) is a direct result of 
non-time-orthogonality. In NTO frames light speed is neither invariant nor 
isotropic. In the time orthogonal frames more typically dealt with in 
special and general relativity, it is invariant and isotropic. Further the 
degree of anisotropy in the speed of light is directly correlated with the 
magnitude of the NTO off diagonal term in the frame's line element.

\subsection*{B. Picturing NTO Frames and Light Speed}

The effect of non-time-orthogonality can be visualized with the aid of 
Figure 3, which shows the time and spatial circumferential axes of both the 
lab frame k and the rotating frame K at radius \textit{r 
}(=\textit{R})\textit{.} We set up our coordinates such that $\phi = z = 
\Phi = Z = 0$, and note that only very small changes in $\phi ,\,\Phi 
,\,t,\,$and \textit{T} are considered. The thinner (orthogonal) lines 
represent the lab frame axes (upper case); the thicker (NTO) lines, the 
rotating frame (lower case).

\begin{figure}[htbp]
\centering
\includegraphics*[bbllx=0.26in,bblly=0.11in,bburx=5.77in,bbury=2.48in,scale=1.00]{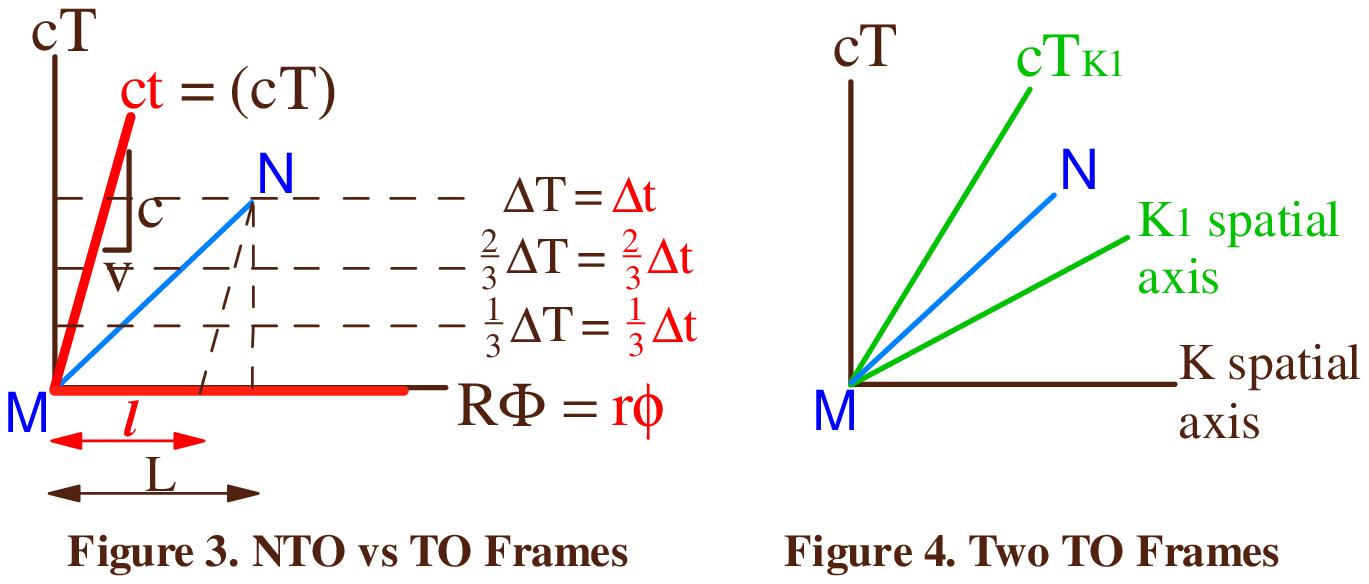}
\end{figure}

\bigskip

Note that the effect of the time dilation factor appearing in $(4)$, 
(5), and (7) is masked, because we are plotting the time axis of the 
rotating frame as our coordinate time \textit{t }(time on a clock at 
\textit{r}=0). This is not the physical time on a standard clock fixed to 
that frame at radius $r \ne 0$. If one wishes, one can simply multiply the 
coordinate time \textit{t} by the (second order) time dilation factor at any 
point in the discussion of this section to obtain exact relationships for 
physical quantities. For the present, however, the primary emphasis is on 
first order effects, as reflected in relationship (1).

From the line element $(4)$ and basic trigonometry (generalized 
Pythagorean theorem) one can determine the slope of the rotating frame time 
axis. Alternatively, set the RHS of (3) equal to RHS of the first line in 
$(4)$ with$d\phi = dr = dz = dR = dZ = 0$, use (2.a) and (2.b), and 
solve for $Rd\Phi /cdT$.

MN is the path of a light ray and has null path length, i.e., \textit{ds} = 
0. Observe that for a given amount of coordinate time (which is the same in 
both k and K, i.e., \textit{c}$\Delta $\textit{T} = \textit{c}$\Delta 
$\textit{t}), the light ray travels a certain spatial distance \textit{l }in 
k, but a greater spatial distance \textit{L} in K. Hence the speed of light 
measured in k is less than that in K, and this corresponds with the plus 
sign before the \textit{c} in (6). For a light ray in the opposite direction 
(minus sign in (6)) one can show graphically (with a light ray in the second 
quadrant of Figure 3 at right angle to MN) that the corresponding \textit{l} 
distance is greater than \textit{L}, and hence the velocity for that ray 
would be greater in k than in K.

Given that the slope of MN is unity, \textit{L = c}$\Delta $\textit{T}. 
Dividing this by $\Delta $\textit{T}, one gets the speed of light in K as 
\textit{c.} The k time axis has slope \textit{c/v = c/}$\omega $\textit{r}, 
so

\begin{equation}
\label{eq8}
l = L - c\Delta T\left( {\omega r/c} \right).
\end{equation}

Dividing this by $\Delta $\textit{T}, one arrives at (6) for the coordinate 
speed of light in k (with the plus sign for \textit{c} since light ray MN is 
traveling in the direction of disk rotation). Note that larger values of 
$\omega $ or \textit{r} mean a greater degree of non-time-orthogonality and 
light speed discrepancy from \textit{c}.

Figure 4, presented for completeness, depicts a co-moving inertial (Lorentz) 
frame K$_{1}$, having instantaneous velocity equal to the circumferential 
velocity of the rotating frame at \textit{r}. Note that in both inertial 
frames K and K$_{1}$, time is (Minkowski spacetime) orthogonal to 3D space 
and the speed of light ray MN equals \textit{c}.

In general one can conclude that non-invariance and degree of anisotropy for 
the local, physically measurable speed of light are directly dependent on 
the slope of the time axis relative to the 3D space axes. All frames for 
which time is orthogonal to space have isotropic light speed equal to 
\textit{c}. All NTO frames have anisotropic light speed not equal to 
\textit{c}.

Note that both the rotating frame and the instantaneously co-moving Lorentz 
frame exhibit some of the same properties. They both have the same time 
dilation as seen from the lab, as well as the same 4D interval \textit{ds} 
between events. However, because of differences in time orthogonality, each 
has different measured speeds for light. Hence, a co-moving Lorentz frame 
can \textit{not} simply be assumed to be an appropriate local surrogate for 
the rotating frame.

\section*{V. Comparison with Experiment}

\subsection*{A. Michelson-Morley Revisited}

Although the analysis herein is supported by the Sagnac experiment, one 
might ask why then did Michelson and Morley not find the speed of light in 
the directions of galactic and solar orbital rotation different from that in 
other directions? The answer, the author submits, is that bodies in 
gravitational orbits follow geodesics, i.e., they are in "free fall". That 
is, they are in locally inertial, time orthogonal frames and therefore obey 
Lorentzian mechanics.

Consider a planet in orbit around a star that doesn't rotate relative to 
distant stars (i.e., one solar day equals one year). An observer inside a 
windowless box (similar to Einstein's enclosed \textit{gedanken} elevator) 
on that planet could do tests (pails of water, Foucault pendulum, Coriolis 
effects, etc.) to determine that she is not rotating. Hence her frame, the 
frame of the planet, would be time orthogonal, and her experiments would 
also find the speed of light to be isotropic, equal to \textit{c}, and 
independent of the planet's orbital velocity.

If, however, her planet were rotating relative to distant stars at $\omega 
$, her measurements would detect a rotation rate of $\omega $, independent 
of orbital angular velocity around her own star. Her frame would then be 
NTO, with all of the concomitant phenomena described in Section IV. These 
phenomena would depend only on $\omega $\textit{r}, the surface velocity of 
the planet relative to the Lorentz frame in which its axis of rotation is 
fixed. No variation in the speed of light would be found from the solar or 
galactic orbital velocities.

It is noteworthy that Michelson and Gale\footnote{ A.A. Michelson, and H.G. 
Gale, ``The effect of the earth's rotation on the velocity of light, Part 
II,'' \textit{Astrophys. J. }\textbf{61}, 140-145 (1925). See also A.A. 
Michelson, "The effect of the earth's rotation on the velocity of light, 
Part I," \textit{Astrophys. J.} \textbf{61}, 137-139 (1925).} measured the 
Sagnac effect for the earth's surface velocity in the 1920's. And in order 
to maintain accuracy, the Global Positioning System must apply a Sagnac 
velocity correction to its electromagnetic signals\footnote{ D. W. Allan, 
and M. A. Weiss, ``Around-the-World Relativistic Sagnac Experiment,'' 
\textit{Science}, \textbf{228}, 69-70 (1985).} .

\subsection*{B. Modern Michelson-Morley Experiment}

Although the original Michelson-Morley experiment and almost all subsequent 
tests of similar nature were not precise enough to detect any non-null 
effect due to the earth surface velocity, one such test was. In 1978, 
Brillet and Hall\footnote{ A. Brillet and J. L. Hall, ``Improved laser test 
of the isotropy of space,'' \textit{Phys. Rev. Lett}., \textbf{42}(9), 
549-552 (1979).} found a "null" effect at the $\Delta $\textit{t}/\textit{t 
}= 3X10$^{-15}$ level, ostensibly verifying standard relativity theory to 
high order. However, to obtain this result they subtracted out a persistent 
``spurious'' non-null signal of amplitude 2X10$^{-13}$ at twice the 
apparatus rotation frequency\footnote{ Mark P. Haugan and Clifford M. Will, 
``Modern tests of special relativity,'' \textit{Phys. Today}, May 1987, 
69-76. It is ironic that Haugan and Will cited the Brillet and Hall 
experiment as proof of the invariance of the speed of light.} .

In 1981 Aspen\footnote{ H. Aspen, ``Laser interferometry experiments on 
light speed anisotropy,'' \textit{Phys. Lett.,} \textbf{85A}(8,9), 411-414 
(1981).} pointed out that this ``spurious'' signal would correspond to a 
test apparatus velocity of 363 m/sec. The earth surface velocity due to its 
rotation at the test site is 355 m/sec. 

\section*{VI. Related Issues}

In this section we briefly discuss several other rotating frame issues that 
are either prevalent in the literature or otherwise worth addressing.

\subsection*{A. Relativistic Mass-Energy}

Klauber\footnote{ Ref. 29, pp. 427-429.} uses transformation (2) to show 
that the mass-energy of an object fixed in the (NTO) rotating frame has 
typical relativistic dependence on speed, in this case $\omega r$. Given the 
well know cyclotron experiments, this lends further support for the 
correctness of transformation (2).

\subsection*{B. Generalized Coordinates and Light Speed}

Some may argue, in the spirit of generalized coordinates common to general 
relativity, that the velocity of light deduced in Section IV is merely a 
coordinate value. That is, it is not what one would measure with standard 
rods and clocks, but an expression in terms of arbitrary coordinates, which 
may in fact be anything one chooses. By choosing the appropriate generalized 
coordinate system we could then ``deduce'' any value we like.

This argument is in fact erroneous. One can calculate physical values for 
distance, time, velocity or any other measurable quantity from the metric of 
the particular coordinate grid employed. (See the Appendix for an example.) 
``Physical'' values are those one would measure using physical instruments 
such as standard rods and standard clocks. ``Coordinate'' values are those 
one calculates using the arbitrary values for length and time associated 
with any arbitrarily chosen grid. In effect, physical values are those 
values one calculates when the associated basis vector of the generalized 
coordinate system has unit length. Given any coordinate value associated 
with a non-unit basis vector, one simply calculates the equivalent 
(physical) value associated with a unit basic vector pointed in the same 
direction. Coordinate values can be any number, while physical values are 
unique. For details we refer the reader to Misner, Thorne, and 
Wheeler\footnote{ Charles W. Misner, Kip S. Thorne, and John A. Wheeler, 
\textit{Gravitation} (Freeman, New York, 1973), pp. 37, 821-822, and many 
other places throughout the text.} , Malvern\footnote{ Lawrence E. Malvern, 
\textit{Introduction to the Mechanics of a Continuous Medium} 
(Prentice-Hall, Englewood Cliffs, New Jersey, 1969), Appendix I, Sec. 5, pp. 
606-613.} , and Klauber\footnote{ Ref. 29, pp. 426-427.} .

Avoiding excessive complexity, we simply note that we have taken care that 
quantities in relationships such as (1) and (7) are indeed physical values. 
We note further that in the thought experiment of section II.A no coordinate 
grid whatsoever is used. The observer simply uses standard rods and clocks, 
which yield the same values regardless of the coordinate grid employed.

A related argument posits that the degree of orthogonality of any axis in 
generalized coordinates relative to any other axis is arbitrary, and hence 
we can choose our time axis in any direction we like. In fact, after using 
transformation (2), some\footnote{ See for example, ref. 30, pp. 124.} 
assert that we must then transform to a locally time orthogonal frame. But 
this is effectively the same as using local Lorentz co-moving frames. This 
in turn necessitates invariant, isotropic light speed and gives rise to 
problems already discussed, as well as the discontinuity in time described 
in the following subsection. While in generalized coordinates we can 
arbitrarily define our \textit{t} coordinate in any of an infinite number of 
ways, if it is to represent the physical time nature chooses, then 
calculations done using it must match up with phenomena observed in the 
physical world.

We also note that any transformation to a rotating frame must incorporate 
the general form of (2.c). That is, the transformation of the azimuthal 
angle must include a term like $\omega $\textit{t}, regardless of whether 
one believes other multiplicative factors (such as the second order Lorentz 
factor) should also be involved. When one squares the \textit{d}$\Phi $ of 
(2.c), or any other relation with a $\omega $\textit{dt }term, and inserts 
the result in (3), one then ends up with off diagonal metric terms such as 
those of $(4)$. Hence no matter what physically reasonable 
transformation\footnote{ For justification of (2.a), see ref. 29, pp. 
416-418.} one chooses, one must find that the rotating frame is NTO. And the 
primary effects of an NTO frame on observable phenomena are first order, 
i.e., they are independent of the presence or absence of a Lorentz factor in 
the transformation.

\subsection*{C. Simultaneity in the Rotating Frame}

Although this and the following subsection may seem counterintuitive to one 
cultured by a relativistic age, we suggest they deserve serious 
consideration as a possible way in which nature might actually work on 
rotating frames.

Using the co-moving local Lorentz frame methodology, one finds, due to the 
standard lack of agreement in simultaneity between Lorentz frames in 
relative motion, a quite bizarre result. (See Klauber\footnote{ Ref. 29, pp. 
Pp. 413-415.} .) If we consider a spatial path 360$^{o}$ around a given 
circumference, we find the clock at 360$^{o}$ has a different time on it 
than the clock at 0$^{o}$, even though time remained constant all along the 
spatial path. This means the clock can not be synchronized with itself. It 
also implies that a continuous standard tape measure laid out around the 
circumference would not meet back up with itself at the same point in time. 
In other words, there would be a discontinuity in time.

Peres\footnote{ Asher Peres, ``Synchronization of clocks in a rotating 
frame,'' \textit{Phys. Rev. D,} \textbf{18}(6), pp. 2173-2174 (1978).} noted 
this result as well, in addition to demonstrating that this methodology led 
to a ``radial velocity of light [that is] not the same inward and outward.'' 
He concluded, ``All this is the heavy price which we are paying to make the 
azimuthal velocity of light ...equal to \textit{c}.''

Reconciliation of analysis with reality occurs if simultaneity on a rotating 
disk is the same as that in the lab. Note that this is true for the time 
transformation of (2.a) wherein \textit{dt = }0 between two events, \textit{ 
}if \textit{dT} = 0 between those same events\textit{.}. Hence (1-$\omega 
^{2}$\textit{r}$^{2}$/\textit{c}$^{2}$)$^{1/2}$\textit{dt}, the time 
passed on standard (physical) clocks in the rotating frame, is also zero. 
(Note that clocks running at different rates can still agree that no time 
passed on either one between events, and so can share a common 
simultaneity.) For this definition of simultaneity, there is no 
discontinuity in time. A line painted around a closed path on the rotating 
disk then does meet back up with itself at the same point in time.

\subsection*{D. Length Contraction: To Be or Not to Be}

The co-moving local Lorentz frames approach implies that standard rods on a 
rotating disk contract circumferentially. Many researchers\footnote{ See 
ref. 3.} have concluded from this that the disk surface is therefore curved, 
thereby ostensibly resolving the famous Ehrenfest paradox\footnote{ P. 
Ehrenfest, "Gleichfömrige Rotation starrer Körper und Relativitätheorie," 
\textit{Phys. Z}. \textbf{10}, 918-918 (1909).} $^{,}$\footnote{ See for 
example, ref. 9.} .

However, Tartaglia's\footnote{ A. Tartaglia, ``Lengths on rotating 
platforms'', \textit{Found. Phys. Lett.}, \textbf{12}(1), 17-28 (1999).} 
interpretation of Ehrenfest's paradox is more insightful. He notes that in 
Lorentz frames each observer sees rods in the other's frame as contracted, 
and ``an observer on board the [rotating] disk would not perceive any 
curvature since in his reference frame [there is no contraction]''.

Still further, if local Lorentz frames are valid surrogates for a rotating 
frame, then an observer on the rim of a rotating disk would likewise see the 
lab rods as contracted. And he would therefore conclude that the lab frame 
must be curved, which of course, it is not. Klauber\footnote{ Ref. 29, pp 
418-419, 431-433, 437, 439.} and Tartaglia both conclude that internal 
contradictions in the theory disappear only if there is no Lorentz 
contraction effect between the disk and lab frames.

Transformation (2) actually implies this. Consider a small circumferentially 
aligned rod of proper length \textit{R}$\Delta \Phi $ ($\Delta \Phi $ is 
small) in the lab, which to a lab observer must have simultaneous endpoints, 
both at time \textit{T}= 0. Using (2.a), (2.b), and (2.c) one then finds the 
length of that rod as seen from the rotating disk to be \textit{r}$\Delta 
\phi $ = \textit{R}$\Delta \Phi $, i.e., no observed Lorentz 
contraction. The same logic works in reverse for a rod on the disk. Again, 
we emphasize that quantities discussed herein are \textit{physical}, not 
merely coordinate, in nature.

The same conclusion may be drawn from the 4D line element \textit{ds}, which 
is invariant between frames, NTO or not, inertial or not. In general between 
any two frames (notation should be obvious)

\begin{equation}
\label{eq9}
ds^{2} = - cdT^{2} + dX^{2} + g_{XT} dXdT = - cdt^{2} + dx^{2} + g_{xt} 
dxdt
\end{equation}

\noindent
where the off diagonal terms vanish for TO frames, and physical quantities 
are assumed (i.e., a coordinate grid with unit basis vectors is chosen.) 
Note that if the two frames share the same simultaneity, then when 
\textit{dT} = 0, \textit{dt} = 0, and therefore \textit{dx} = \textit{dX}. 
The length \textit{dx }of a rod in one frame equals the length \textit{dX} 
of the same rod as seen from the other frame, i.e., there is no Lorentz 
contraction. (Note that if \textit{dT }= 0, but \textit{dt} $ \ne $ 0, then 
\textit{dx} $ \ne $ \textit{dX}, and there is Lorentz contraction.\footnote{ 
There is a little subtlety here. Calculation of the Lorentz contraction 
actually involves rod endpoints that appear to be different events to 
different observers. The present example assumes the same two events are 
measured by both observers. However, the conclusion remains valid. If the 
endpoint events look simultaneous to two different observers, then the rod 
length measured must be the same for each.} ) 

The Lorentz contraction results from the Lorentz transformation, and in a 
sense is little more than an optical illusion. No Lorentz contracted object 
ever ``feels'' contracted\footnote{ See ref. 29, pp. 415, 418-419, 423.} . 
The contraction appears because of the disagreement in simultaneity between 
frames, which is inherent within the Lorentz transformation. 

We conclude that if two frames share the same simultaneity, then there is no 
Lorentz contraction effect between those frames. Hence, equivalence of 
simultaneity for the lab and rotating frames means no Ehrenfest paradox, as 
well as no discontinuity in time\footnote{ Neither Lorentz contraction nor 
simultaneity differences can presently be measured directly by experiment.} 
.

\subsection*{E. Absolute Nature of Rotation}

Translating systems display robust relativity. For such systems, absolute 
velocity does not exist, and there is no way to determine a preferred frame. 
As a result, light speed is invariant and transformations must be 
Lorentzian.

Rotating systems differ from translating systems in that one can determine 
one's angular velocity absolutely (in the sense of Mach). The preferred 
frame is then the non-rotating frame, which any observer can readily 
identify. Further, an observer within her own frame can, in fact, do tests 
that unambiguously determine her circumferential velocity relative to the 
Lorentz frame in which her axis of rotation is fixed. 

If two types of frames have such fundamental dissimilarity at their cores, 
is it not presumptuous to assume, as has been the general practice, that the 
phenomena associated with each are identical? Is it correct to simply 
presume that invariant light speed, Lorentz contraction, and disagreement in 
simultaneity can be directly extrapolated to rotating frames? We suggest 
that it is not, and that the proper course of action consists of building a 
theory of relativistic rotation from empirical data on rotating frames 
alone, independently of preconceived conceptions.

\section*{VII. Summary and Conclusions}

The analysis of the previous section may, of course, run counter to a few 
long cherished ideas. But it does cleanly resolve certain major issues, and 
in the process leaves the essence of relativity theory intact. 

Invariants like \textit{ds} remain invariant. Every aspect of the theory for 
time orthogonal frames (the vast majority of applications) remains 
unchanged. Lorentz frames are still related by Lorentz transformations (with 
concomitant effects such as Lorentz contraction, etc.), and differential 
geometry continues its reign as descriptor of non-inertial systems. Neither 
special nor general relativity need be altered in any regard, provided of 
course, that NTO frames are appropriately interpreted.

All of the results obtained herein are derived from two postulates: i) 
transformation (2) relates rotating and non-rotating frames, and ii) the 4D 
line element \textit{ds} is invariant. If these postulates are valid, it 
appears one must conclude the following.

NTO frames display non-invariant and non-isotropic local, physical speed of 
light, to a degree dependent on the degree of non-time-orthogonality. 
Lorentz frames are appropriate local surrogates for (curved or flat) TO 
frames, but not for (curved or flat) NTO frames. Rotating frames are truly 
(not superficially) NTO, and if one makes a straightforward interpretation, 
unfettered by preconceptions, of the most widely accepted transformation, 
one concludes that rotating and non-rotating frames share the same 
simultaneity. This in turn implies an absence of the Lorentz contraction 
effect between those frames.

NTO analysis for rotating frames predicts time dilation and mass energy 
increase with tangential speed in consonance with cyclotron experiments. It 
is also in full agreement with the Sagnac experiment and related thought 
experiments. Importantly, it also explains the persistent signal found in 
the Michelson-Morley type Brillet and Hall experiment, which has heretofore 
been considered inexplicable.

\appendix
\section{Appendix}

The following discussion should be read only in the context of time 
orthogonal reference frames. Such frames make up the bulk of all 
applications, and virtually 100\% of textbook problems. The conclusions 
drawn in this appendix are subsequently modified for NTO frames.

The speed of light in non-Lorentzian systems can be a source of confusion as 
it is sometimes (misleadingly) said that the speed of light in general 
relativity can be different than \textit{c}. This is true if, for example, 
one measures the speed of light near a massive star using a clock based on 
earth. (Time on such a clock is effectively the coordinate time in a 
Schwarzchild coordinate system.) As is well known, due to the intense 
gravitation field, the passage of time close to the star is dilated relative 
to earth time, and using the earth clock, one would indeed calculate a light 
speed other than \textit{c}. However, use of standard rods and clocks 
adjacent the light ray itself would result in a speed of precisely 
\textit{c}. In the language of Section VI.B, the speed of light calculated 
with the earth clock is a \textit{coordinate} speed, whereas that measured 
with rods and clocks proximate to the light ray is the \textit{physical} 
speed.

Other confusion exists for scenarios where spacetime itself expands or 
contracts. For example, just after the big bang, space itself was expanding 
much like the surface of a balloon being blown up. A photon in space 
(analogous to an ant on the surface of the balloon) at a different location 
than an observer could then move away from the observer faster than 
\textit{c} (analogous to faster than the ant can crawl on the surface) 
because the space (balloon surface) between the photon and the observer is 
itself expanding. Yet a photon spatially coincident with an observer could 
never be seen by that observer to have speed greater than \textit{c}, and 
local standard rods and clocks adjacent any photon would find its speed 
equaling \textit{c} regardless of the dynamical state of spacetime itself.

More mathematically, for a given generalized coordinate system in a 
non-inertial frame, we have

\[
ds^{2} = g_{tt} c^{2}dt^{2} + g_{xx} dx^{2} + g_{yy} dy^{2} + g_{zz} dz^{2}
\]

\noindent
where \textit{g}$_{tt}$ is negative. Consider a ray of light passing in the 
\textit{x} direction such that \textit{dy = dz =} 0. The coordinate speed of 
light is found by setting \textit{ds} = 0 and solving for the length in 
generalized coordinates (the coordinate length) that the light ray travels 
divided by the time in generalized coordinates (the coordinate time), i.e.,

\[
\frac{{dx}}{{dt}} = \sqrt {\frac{{ - g_{tt} }}{{g_{xx} }}} \;c.
\]

On the other hand, the physical speed of light is the physical length 
divided by the physical time. Physical length (measured by standard rods) is 
$\sqrt {g_{xx} } dx$\footnote{ See refs. 38,39, and 40.} , and physical time 
(measured by standard clocks) is $\sqrt { - g_{tt} } dt$. So the speed of 
light as measured by physical instruments, regardless of the generalized 
coordinates chosen, is

\[
\frac{{\sqrt {g_{xx} } dx}}{{\sqrt { - g_{tt} } dt}} = \;c,
\]

\noindent
and this is always equal to \textit{c} (for time orthogonal frames.)

\end{document}